\documentstyle[sprocl]{article}
\input{psfig}
\def\Journal#1#2#3#4{{#1} {\bf #2}, #3 (#4)}

\def\NPB{{\em Nucl.~Phys.}~B}
\def\PLB{{\em Phys.~Lett.}~B}
\def\PRL{\em Phys.~Rev.~Lett.}
\def\PRD{{\em Phys.~Rev.}~D}

\def\be{\begin{equation}}
\def\ee{\end{equation}}
\def\bea{\begin{eqnarray}}
\def\eea{\end{eqnarray}}
\def\VEV#1{\langle #1\rangle}
\def\gsim{\mbox{\raisebox{-.6ex}{~$\stackrel{>}{\sim}$~}}}

\begin{document}

\title{ELECTROWEAK PHASE TRANSITION AND BARYOGENESIS\\ IN
 THE MINIMAL SUPERSYMMETRIC STANDARD MODEL}

\author{J.M.~CLINE}

\address{McGill University, Department of Physics,
Montr\'eal, Qu\'ebec H3A 2T8, Canada}

\maketitle\abstracts{I describe work done in collaboration with
M.~Joyce and K.~Kainulainen on (1) the strength of the electroweak
phase transition in the MSSM and (2) the mechanism for producing the
baryon asymmetry during the phase transition.  In the former we compare
the effective potential and dimensional reduction methods for
describing the phase transition and search the parameter space of the
MSSM for those values where it is strong enough.  In the latter we give
a systematic computation of the baryon asymmetry due the CP-violating
force acting on charginos in the vicinity of the bubble wall.  We find
that a light right-handed stop, a light Higgs boson, and a large phase
in the $\mu$ parameter, are the main necessary ingredients for
producing the baryon asymmetry.}

\section{Experimentally testable baryogenesis?} A major triumph of
astroparticle physics is our understanding of the light element
abundances through big bang nucleosynthesis.  Equally exciting is the
prospect of putting the measured baryon asymmetry of the universe,
$\eta = n_B/n_\gamma\cong 3\times 10^{-10}$, on a similar footing.
Unfortunately, there is much less agreement about the mechanism of
baryogenesis than there is for nucleosynthesis.  A search of the SPIRES
database shows about 400 papers on the subject so far, 350 of which (my
rough estimate) describe different ways of generating the baryon
asymmetry.  Which one of them, if any, is correct?  Of course it
is possible that the asymmetry was generated at such high temperatures
that direct confirmation of the new physics needed will never be
achieved in the laboratory.  In this case baryogenesis should forever
be relegated to the realm of untestable speculation, which is not very
interesting from the point of view of phenomenology.

More interesting are the scenarios which can be falsified
experimentally before we retire.  Electroweak baryogenesis is arguably
the most attractive such possibility at present.  It needs new physics
at the 100 GeV energy scale, of a kind that is even now being probed by
the LEP 2 experiment.  It thus holds out the hope that it could be ruled
out in the near future, reducing the number of candidate baryogenesis
mechanisms from 350 to 349: this is progress!  On the other hand, if it
is not ruled out we may be on the way to a truly quantitative theory of
the origin of baryonic matter, with consequences more profound than
nucleosythesis for our understanding of particle physics and the early
universe.

The basic picture of how electroweak baryogenesis could work was
established by Cohen, Kaplan and Nelson\cite{CKN} in 1992.  The
electroweak phase transition is supposed to be first order, meaning
that at some critical temperature $T\sim 100$ GeV, the bubbles of the
broken phase (where the Higgs field VEV is nonzero) begin to nucleate
in the high temperature symmetric phase (where $\VEV{H}=0$).  The
bubbles expand, collide and eventually fill up the universe with the
broken phase.  Before doing so however, particles in the high-$T$
plasma will be partially reflected when encountering the bubble walls,
as can be understood from the fact that they gain mass from the nonzero
Higgs VEV as they enter the bubbles.  These interactions with the
bubble walls can be CP violating.  For example, left-handed top quarks
may reflect from the bubble walls more strongly than right-handed top
quarks, leading to a buildup of chirality in front of the expanding
bubble wall.  

\begin{figure}
\centerline{\psfig{figure=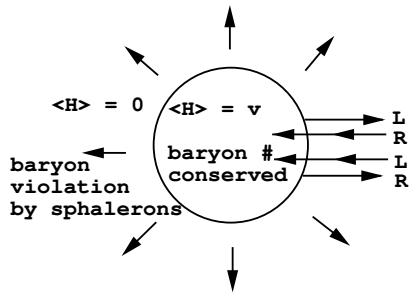,height=1.5in}}
\caption{Expanding bubble during the electroweak phase
transition.  Left-handed fermions reflect into right-handed ones
and vice versa, but with different probabilities.}
\end{figure}

Enter the sphaleron, the electroweak field configurations which violate
baryon number in the Standard Model due to the fact that the baryon
current is anomalous in the presence of background $W$ boson fields.
Each sphaleron interaction creates or destroys 9 left-handed quarks and
3 left-handed leptons in such a way as to violate $B$ (baryon number)
and $L$ (lepton number) by 3 units, while preserving $B-L$.
Sphalerons, ``seeing'' the excess of chirality in front of the wall,
will tend to drive it to zero by converting excess left-handed quarks
into their antiparticles.  However this in turn creates a baryon
asymmetry.  Eventually the excess baryons in front of the wall are
overtaken by the wall and fall inside the bubbles, where the sphalerons
are much heavier than they are in the symmetric phase.  If  the
sphalerons are sufficiently Boltzmann-suppressed inside the bubbles,
the baryon excess is safe from erasure by sphalerons, which would
otherwise, given enough time, tend to cause it to relax to zero.  (The
competing tendency to erase chirality is no longer present inside the
bubbles because the particles have masses there, and so chirality is
not a good quantum number.)

New physics is needed by electroweak baryogenesis for the following
reasons.  (1) The phase transition should be strongly enough first
order to prevent the sphalerons from erasing the baryon asymmetry
inside the bubbles, and (2) the CP violation on the wall must be strong
enough to generate enough baryons.  A natural candidate for the origin
of such new physics is the minimal supersymmetric standard model
(MSSM), since it is able to fulfill both of these criteria.\cite{ewb}

\section{Strength of the phase transition: when is ${\bf v_c/T_c >1}$?}

To avoid the washout of baryons inside the bubbles once they have been
created, the sphaleron configuration must be heavy compared to the
temperature.  In this case the rate of sphaleron interactions inside
the bubbles goes like $\Gamma_{\rm sph} \sim e^{-{\rm const.}
(v_c/T_c)}$ where $v_c$ is the Higgs field VEV (normalized so that
$v=246$ GeV at $T=0$) in the bubbles at the critical temperature $T_c$,
when nucleation takes place (figure 2).  This follows from the fact
that the sphaleron energy is proportional to $v_c$.  One can show that
baryon preservation requires that $v_c/T_c$ be greater than 1.0$-$1.1,
the exact number depending on how the potential evolves with
temperature.  In the standard model it is impossible to satisfy the
$v_c/T_c>1$ constraint, but in the MSSM it has been shown that the
right-handed top squarks, if sufficiently light, can enhance the
strength of the phase transition to the required level.

\begin{figure} 
\centerline{\psfig{figure=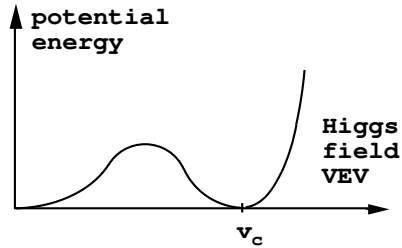,width=2.05in}}
\caption{Schematic picture of the effective potential of the Higgs
field near the critical temperature.} 
\end{figure}

I will review two methods that have been used to compute the critical
ratio $v_c/T_c$.  Traditionally one used the finite-temperature
effective potential (EP) at one or two loops.\cite{espinosa}\ \ However
at high $T$ the perturbative expansion is no longer controlled by the
usual dimensionless couplings, because it is effectively a 3D theory
with potential infrared divergences from the light (transverse $W$
boson) degrees of freedom whose thermal masses are small.  These IR
problems can render the predictions of the EP unreliable.  Roughly
speaking, this happens if the ratio $\lambda T_c/m\gsim 1$, where $m$
is the thermal mass of the light particle at $T_c$ and $\lambda$ is its
quartic coupling to the Higgs boson.

A possible way around this problem is dimensional reduction
(DR),\cite{DR} where one computes an effective theory by integrating
out all {\it but} these most dangerous infrared contributions, leaving
the lattice to take care of them.   In DR, an effective three
dimensional Lagrangian is obtained which has the form ${\cal L}_{\rm
eff} = |\vec D H|^2 + F_{ij}F_{ij}/4g_3^2 -\mu^2_3|H|^2 +
\lambda_3|H|^4$, where $H$ is the light linear combination of the two
Higgs doublets of the MSSM.  It has been found\cite{KLRS} that the limit
$v_c/T_c>1$ corresponds to $x_c\equiv\lambda_3/g_3^2< 0.044$.  Thus one
need only compute $x_c$ in terms of the parameters of the MSSM to find
which ones are consistent with baryogenesis.

We have searched the MSSM parameter space using both the EP and DR at
one loop to compare the two approaches,\cite{ck} and found that they
are roughly compatible, although DR gives a somewhat bigger range of
allowed parameters, as shown in figure 1.  The most sensitive
parameters along with their predicted values are $\tan\beta\equiv
\VEV{H_2}/\VEV{H_1} = 2^{+1.5}_{-0.5}$, $m_{h^0} < 85$ GeV, a
light stop with $m_{\tilde t_1} = 150-200$ GeV, and a correspondingly
small right-handed soft-breaking stop mass parameter of $m_U< 100$
GeV.  Negative values of $m_U^2$ have been advocated for making the
transition strong enough, but our results show that although they are
helpful, they are not required.  Furthermore we find that arbitrarily
large values of $\mu$ are allowed, which is encouraging because $\mu$
is the major source of CP violation in the model.  This is in contrast
to previous studies that have emphasized the weakening effect of $\mu$
on the transition.  We will comment upon the phenomenological relevance
of these parameter values below.

\begin{figure}
\centerline{\psfig{figure=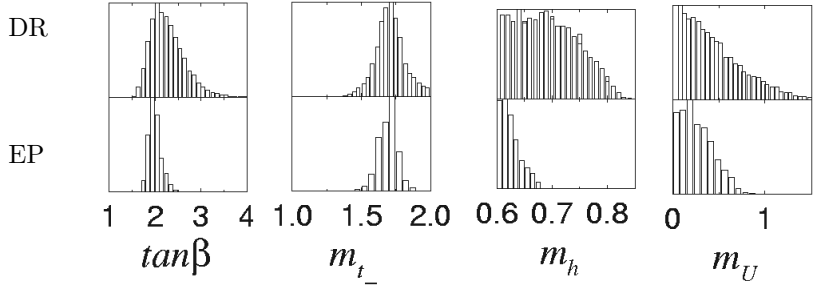,height=1.5in}}
\caption{Distributions of parameters satisfying the baryogenesis phase 
transition
constraint in the MSSM, using dimensional reduction (top line) and the
effective potential (bottom line).  Masses are in units of 100 GeV.}
\vskip-2in
\leftline{DR}
\vskip0.5in
\leftline{EP}
\vskip 1in
\end{figure}

It should be mentioned that studies of the EP at two loops show a
less restrictive range of allowed parameters, more in agreement with
DR at one loop.\cite{espinosa}\ \ It would be interesting to extend the
DR analysis to two loops to see if agreement between the two methods 
is further improved.

\section{Origin of the baryon asymmetry---classical force mechanism}
Knowing the preferred parameters for safeguarding the baryon asymmetry,
we now ask whether it can be generated in sufficient quantity in the
first place.  At first sight this might appear to be a very daunting
problem, because the particles reflecting off the bubble wall are
colliding with other particles in the plasma while they are quantum
mechanically reflecting from the potential of the bubble wall.  This
mixture of particle diffusion and quantum reflection might compel us to
use the quantum Boltzmann equation, a very unwieldy formalism.

However the bubble wall has a thickness of $L\sim 15/T$, while the
average particle in the plasma has a DeBroglie wavelength of
$\lambda\sim 0.3/T$.  This means that its interaction with the wall
is not really quantum mechanical (such effects are exponentially
suppressed when $\lambda \ll L$), but rather can be derived from a
classical force $F$ exerted by the wall on the particles.\cite{JPT}\ \
We are thus justified in using the usual Boltzmann equation, $\partial
f/\partial t - \vec v\!\cdot\!\vec\nabla_x f + \vec
F\!\cdot\!\vec\nabla_p f = {\cal C}[f]$.  The force $F$ acts with
opposite sign on opposite chiralities, leading to chiral charge
separation in front of the wall.  This can be derived in a rigorous
way, starting from a WKB solution to the particles' equation of motion
in the vicinity of the wall.\cite{cjk}

We find that the most important contribution to the baryon 
asymmetry is from Winos and Higgsinos reflecting from the
wall and subsequently being converted (via top quark Yukawa interactions)
into a quark asymmetry that biases
the sphalerons.  The CP violation appears in the chargino mass matrix
through the phase of the $\mu$ parameter:\newline
\centerline{
$\left(\begin{array}{cc} \overline{\tilde W}_R^+ & 
	\overline{\tilde h}^+_{1,R} \end{array}\right)
\left(\begin{array}{cc}  m_2 & g v_2(z)/\sqrt{2}\\
	g v_1(z)/\sqrt{2} & |\mu|e^{i\delta} \end{array} \right)
	\left(\begin{array}{c} \tilde W_L^+ \\ \tilde h^+_{2,L}
	\end{array}\right)$.}

The final answer for the baryon asymmetry depends on $m_2$ (the
Wino mass parameter), $\mu$, $\tan\beta$, the Higgs and quark diffusion
constants, the wall velocity $v_w$, the strong and weak sphaleron
rates, the rate of top Yukawa interactions, the wall width, and the
critical VEV and temperature. It is also proportional to the underlying
CP violation in the $\mu$ parameter, $\sin\delta$.  Putting in our best
estimates for these quantities (or the ranges allowed by the constraint
$v_c/T_c>1$) we find that typically $\sin\delta$ must be unity to get a
large enough asymmetry, although if $m_2\cong\mu$ then it is
possible to have $\sin\delta \sim 0.1$.  The dependence of
$\eta_{10}=10^{10}n_B/n_\gamma$ on $\mu$ and $m_2$ is shown for
several choices of other parameters in figure 2.

\begin{figure}
\centerline{\psfig{figure=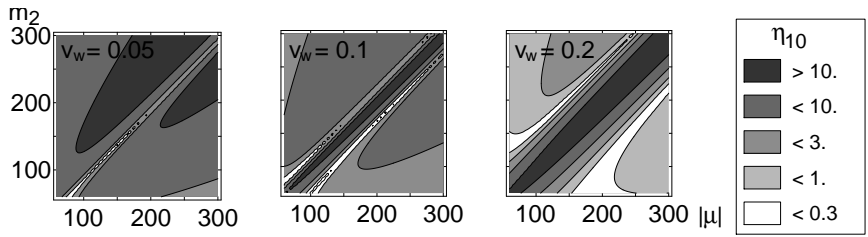,height=1.21in}}
\caption{Contours of constant $\eta_{10}=10^{10}n_B/n_\gamma$ in the
plane of $m_2$ (the Wino mass) and $\mu$, in GeV, $v_{\rm wall}=
0.05$, $0.1$ and $0.2$, $\tan\beta=2$, $\sin\delta=1$ and wall width $=
14/T$.} 
\end{figure}

\section{Relation to Collider Physics and EDM Searches}

We have shown that electroweak baryogenesis in the MSSM works only for
rather limited ranges of parameters like $2.5< \tan\beta< 4$ and the
lightest higgs mass $m_{h^0}<85$ GeV.  In fact, the latter provides the
strongest prospect for falsifying the theory since, as shown in figure
5,\cite{yellow}\ 
 $m_{h^0}$ up to 90 GeV should be ruled out or discovered by the
completion of LEP 2.   Moreover, values of $\tan\beta$ up to 4 will be
explored, which is essentially the same thing since the mass of the
light Higgs is related to $\tan\beta$ in a simple way.

\begin{figure}
\centerline{\psfig{figure=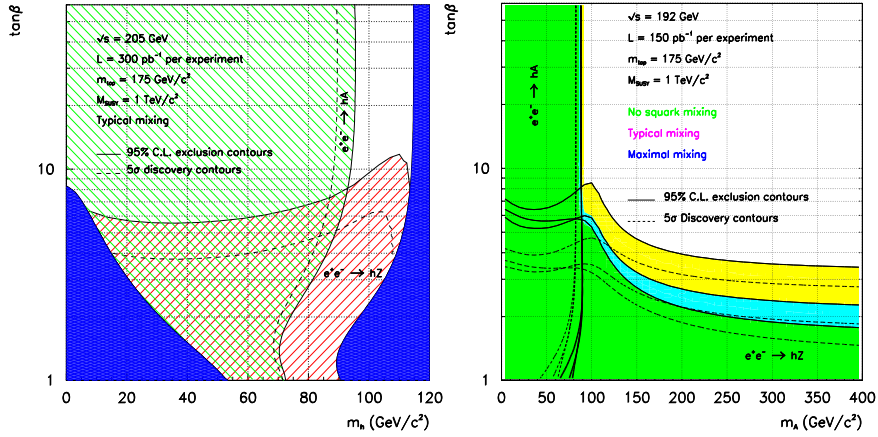,height=2.25in}}
\caption{LEP2 discovery potential for lightest Higgs boson 
in MSSM, and sensitivity to $\tan\beta$. Courtesy of P.~Janot (ALEPH).}
\end{figure}

\begin{figure} 
\centerline{\psfig{figure=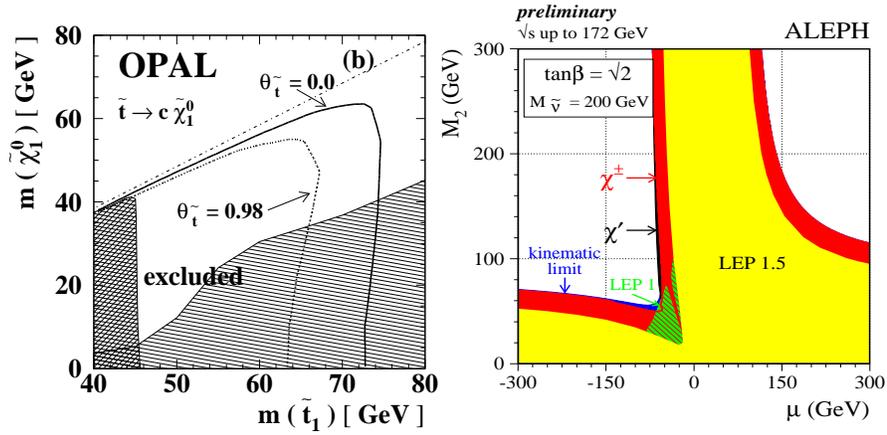,height=2.25in}}
\caption{Exclusion plots for (a) neutralinos versus light stops,
depending on stop mixing angle $\theta_t$, and (b) $m_2$ versus $\mu$.
Double-hatched region in (a) is excluded by D0 at Tevatron.  Courtesy
of J.-F.~Grivaz$^{10}$ (ALEPH). } 
\end{figure}

A less robust signal of new physics is the relatively light top squark
required.  Figure 6a \cite{grivaz} shows how the sensitivity to the
one-loop decay channel $\tilde t\to c\tilde\chi^0_1$ (charm plus
neutralino) depends strongly on the neutralino mass.  Similarly figure
6b shows that much of the $\mu-m_2$ plane relevant for baryogensis is
already excluded by LEP 1.5, but under the assumption of heavy
sneutrinos (pair production of charginos is reduced by destructive
interference from $\tilde\nu$ exchange if the latter are light).

A potentially serious challenge to the naturalness of baryogenesis in
the MSSM is the very large value indicated for the $\mu$ parameter
phase $\delta$.  Values of $\delta>10^{-3}$ would normally be
considered as incompatible with bounds on the neutron electric dipole
moment, but this is only true if the lower generation squarks and the
charginos are all relatively light.  If the up and down squarks are
much heavier than the stop and sbottom, say 1 TeV, then the EDM bound
on $\delta$ is relaxed.  One might object that it looks unnatural to
make a hierarchy between the first two generations and the third
generation squarks.  It is encouraging that such a scenario has
recently been proposed for completely different reasons.\cite{Eff_susy}

\section{Summary} It appears that producing the baryon asymmetry in the
MSSM is not easy; the parameter space is highly constrained.  From the
perspective seeking an experimentally falsifiable theory of
baryogenesis, this is a encouraging.  The phase transition is strong
enough only with a relatively light stop and higgs, and $\tan\beta\sim
2$.  Baryon production is maximized when $\mu\sim m_2$, hence the
Wino and Higgsino are roughly degenerate.  Also the CP violation in the
$\mu$ parameter must satisfy $\sin\delta\gsim 0.1$, which indicates a
large neutron EDM, in fact too large unless the lower generation
squarks are much heavier than the light stop.  If supersymmetry should
be confirmed in the next few years and these predictions verified, it
will be very tempting to believe we finally have evidence for the true
origin of the baryon asymmetry of the universe.

\end{document}